\documentclass{iaus}
\usepackage{graphicx}

\title[MHD of turbulent core collapse] 
{3D MHD model of the collapse and fragmentation of turbulent prestellar core}

\author[Alexander Dudorov \& Sergey Zamozdra]
{Alexander E. Dudorov, \and Sergey N. Zamozdra}

\affiliation{Chelyabinsk State University, Russia \\ email: {\tt dudorov@csu.ru, sezam@csu.ru} }

\pubyear{2010}
\volume{270}
\pagerange{???--???}

\jname{Computational Star Formation}
\editors{J. Alves, B. Elmegreen, J. Girart \& V. Trimble, eds.}

\begin{document}

\maketitle

\begin{abstract}
With the help of 3D MHD simulations we investigate the collapse and fragmentation 
of rotating turbulent prestellar core embedded into turbulent medium. 
The numerical code is based on a high resolution Godunov-type finite-difference 
scheme. Initial turbulence is represented by the ensemble of Alfven waves with 
power law spectrum. Our computations show that under realistic parameters two 
bound fragments can appear when the density increases at $100-10^3$ times. The 
distance between the fragments is about $0.1$ of the initial core radius and their 
orbital period is comparable to the initial free fall time of the core. These 
results can explain the origin of binary stars with separation $0.001-0.01$ pc
in the Galaxy field.
\end{abstract}

\section{Introduction}
Prestellar cores in molecular clouds are gravitationally bound condensations with 
stellar masses. The collapse and fragmentation of such cores lead to formation 
of single or multiple stars. We investigate the influence of magnetic field, rotation 
and turbulence on a prestellar core collapse and early fragmentation using 3D 
magnetohydrodynamical (MHD) simulations. Similar problem without turbulence was 
studied by \cite[Machida et al. (2008)]{Machida08}, 
\cite[Hennebelle \& Teyssier (2008)]{Hennebelle08} and 
\cite[Duffin \& Pudritz (2009)]{Duffin09}. The role of turbulence was investigated by 
\cite[Price \& Bate (2008)]{PriceBate08} and \cite[Wang et al. (2010)]{Wang10} for 
massive clouds without rotation.

We carry out low resolution simulations (with cell number 128$^3$) and focus the 
attention on initial conditions and results analysis. The computational domain 
contains both the core and surrounding medium to describe a turbulent energy 
redistribution via MHD waves propagation (see \cite[Heitsch et al. 2001]{Heitsch01}). 
The initial pulsations amplitudes depend on density (see \cite[McKee \& Zweibel 1995]{MZ95})
because a turbulence in non-uniform medium is inhomogeneous one. A bound fragment 
is found with the help of internal motion separation on inward, outward and 
tangential components relative to fragment mass center.

\section{Model}
\begin{figure}[b]
\begin{center}
 \includegraphics[width=3.4in, trim=0.8in 0.25in 0.8in 0.3in, clip]{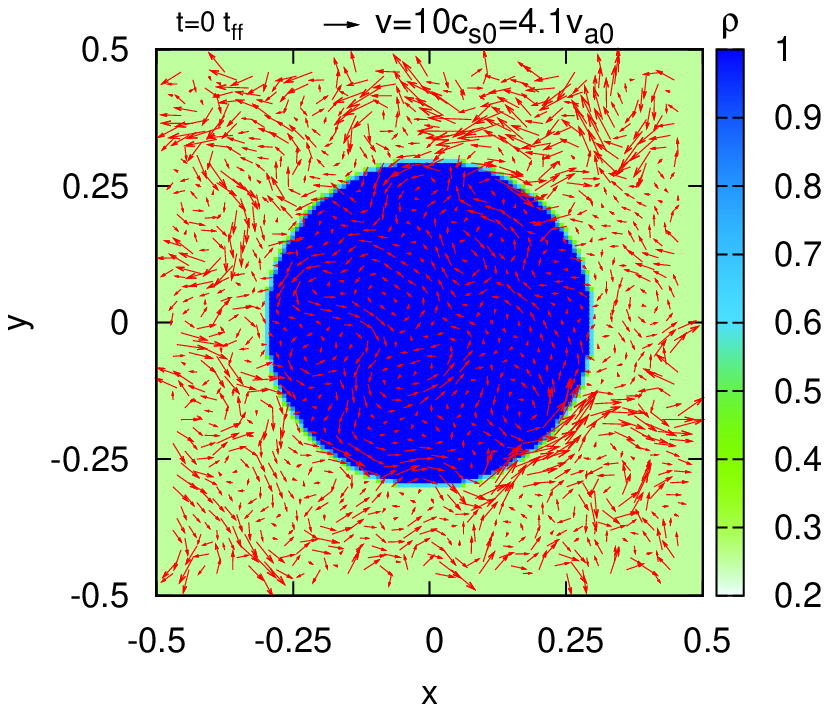}
 \includegraphics[width=3.4in, trim=0.8in 0.25in 0.8in 0.3in, clip]{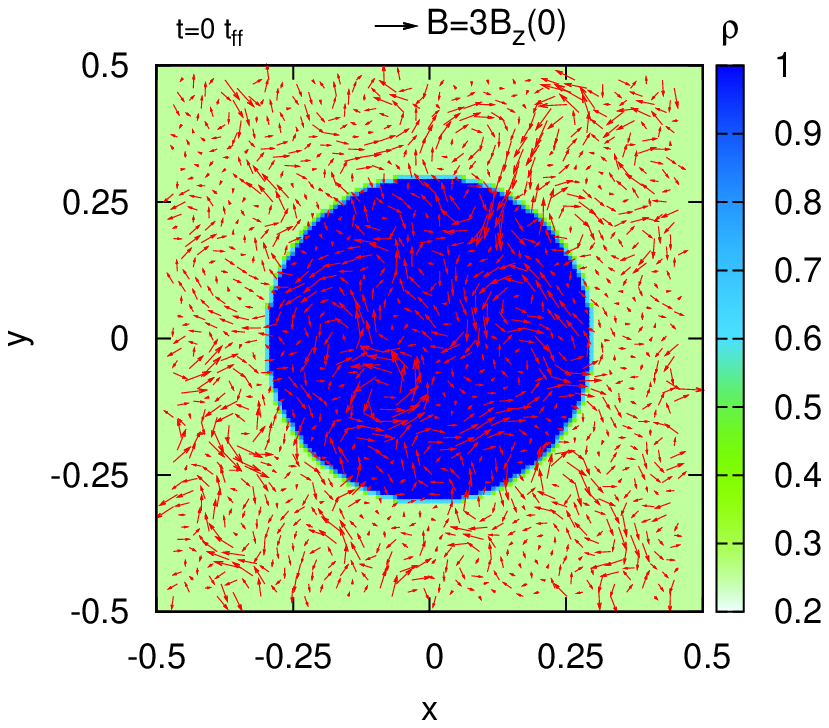}
 \includegraphics[width=3.4in, trim=0.8in 0.25in 0.8in 0.3in, clip]{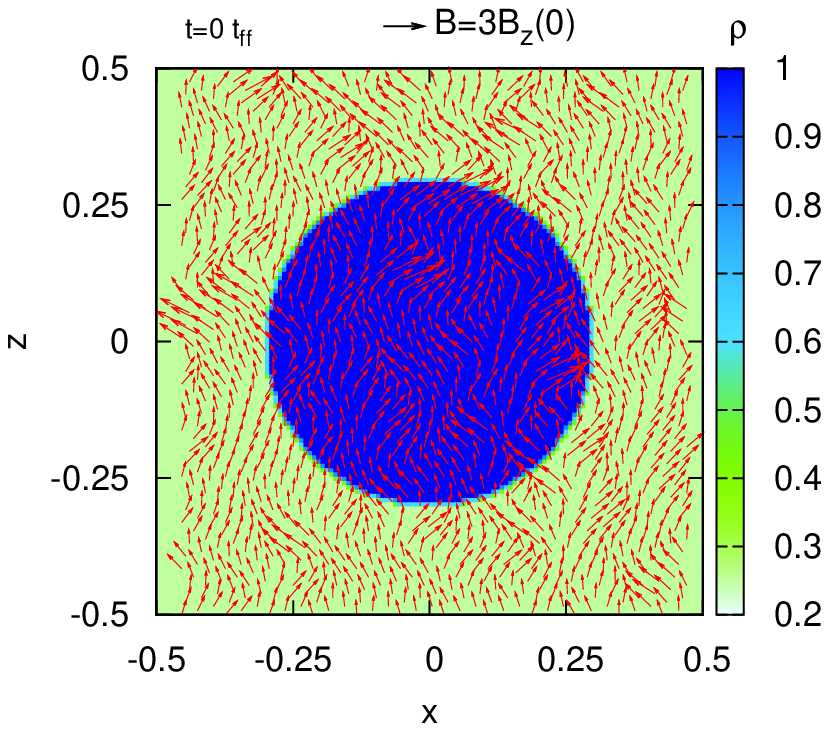} 
 \caption{Initial distributions of density (color scale), velocity and magnetic field 
 (arrows) at different planes in the base model.}
 \label{f:rhovB_ini}
\end{center}
\end{figure}

{\underline{\it Initial conditions}}. Uniform spherical core with radius 
$R=0.3$ of domain size is embedded into uniform cubic medium (see fig. 
\ref{f:rhovB_ini}) with density $\rho=1/4$ of core density. Thermal pressure
is balanced. Large scale magnetic field $\vec B_0$ is uniform. The core and medium 
can rotate as solid body with axis parallel to $\vec B_0$. Turbulence is the ensemble 
of plane Alfven waves with linear polarization and isotropic power law spectrum. The 
initial velocity pulsation $\vec v_{\vec k}$ in the Alfven wave with wave vector 
$\vec k$ is directed perpendicularly to $(\vec k,\vec B_0)$ plane and has amplitude
\begin{equation}
  v_{\vec k}(\vec r) = A k^{-\alpha} \rho^{\beta}(\vec r) 
  \sin\left( \vec k \vec r + \varphi_{\vec k} \right) \,,
\end{equation}
where $A$ is the normalization factor, $\varphi_{\vec k}$ is the random phase.
The magnetic pulsation is $\vec B_{\vec k}=\mp s\vec v_{\vec k}\sqrt{\rho}$, where
$s=sign(\vec k\vec B_0)$.

{\underline{\it Boundary conditions}}. Periodical boundary conditions are used
to promote the MHD waves circulation through the medium.

{\underline{\it Equations and methods}}. Ideal MHD approximation with isothermal 
equation of state is adopted. The simulations are carried out with the help of 
numerical code Megalion \cite[(Dudorov et al. 2003)]{Dudorov03} that realizes the 
modified Lax-Friedrichs method for hyperbolic equations and conjugated gradients 
method for gravitation equation. The computations are single threading.

{\underline{\it Bound fragments}}. A fragment is the set of contacted 
cells where density exceeds some level. The fragment is bound if
\begin{equation}
  E_{g} + E_{in} > E_{th} + E_{m} + E_{tan} + E_{out} \,,
\end{equation}
where $E_g$ is the absolute value of gravitational energy, $E_{th}$, $E_m$ are the 
thermal and magnetic energies, $E_{in}$, $E_{out}$, $E_{tan}$ are the energies 
of inward, outward and tangential motions relative to fragment mass center.

{\underline{\it Basic parameters}}.

1) The ratios of thermal $E_T$, large scale magnetic $E_M$, turbulent kinetic $E_K$ and 
rotational $E_\Omega$ energies of the core to the absolute value of its 
gravitational energy $E_G$.

2) The index of wave spectrum: $\alpha \in(1/4, 1/2)$.

3) The index of $v_{\vec k}-\rho$ relation: $\beta\in(-1/2, 1/4)$. Note that $\beta=-1/2$ 
means the balance of wave pressure.

4) Maximal and minimal dimensionless wave numbers: $n_{min}\in(2,3)$ and 
$n_{max}\in(8,12)$.

In all the models $E_T/E_G=0.1$. The model with parameters $E_\Omega/E_G=0.05$, 
$E_K/E_G=0.3$, $E_M/E_G=0.3$, $\alpha=1/3$, $\beta=-1/2$, $n_{min}=2$ and 
$n_{max}=8$ is called "base model".

\section{Results}
\begin{figure}[b]
\begin{center}
 \includegraphics[width=3.4in, trim=0.8in 0.25in 0.8in 0.3in, clip]{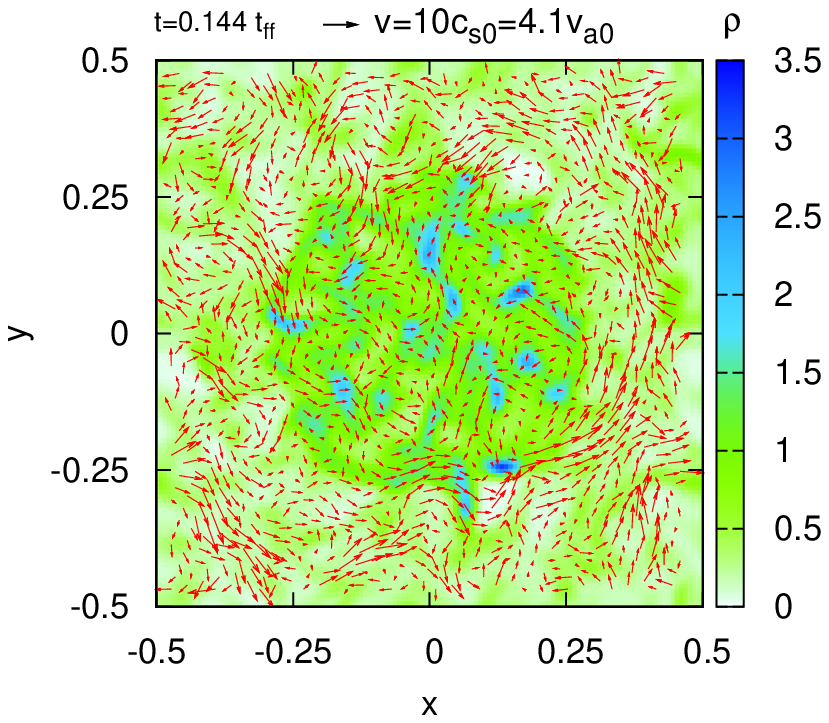}
 \includegraphics[width=3.4in, trim=0.8in 0.25in 0.8in 0.3in, clip]{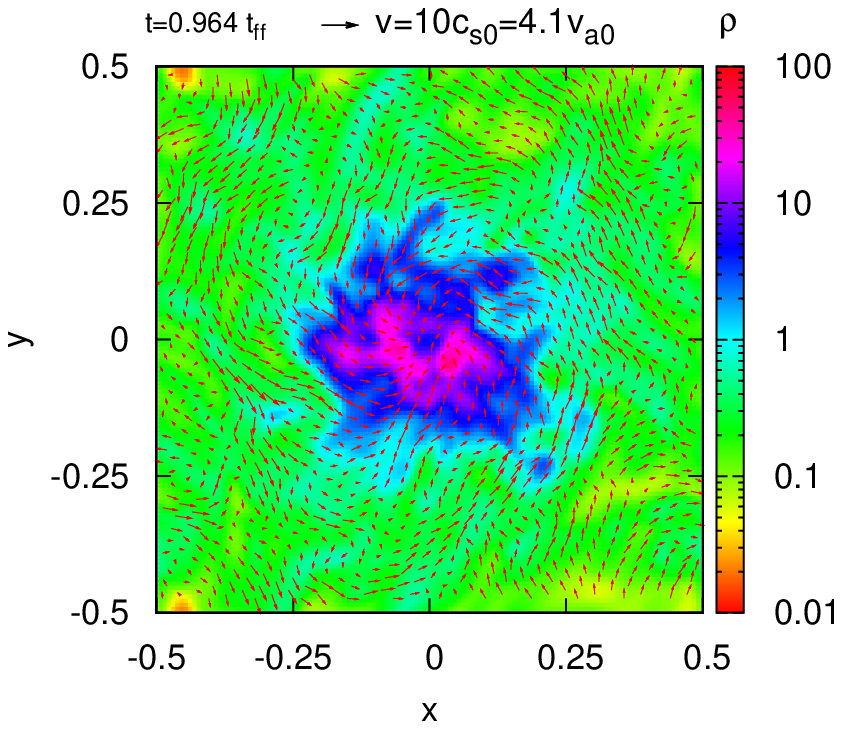}
 \caption{Distributions of density (color scale) and velocity (arrows) at the plane 
 $z=0$ in the base model at $t=0.144 t_{ff}$ and $t=0.964 t_{ff}$.}
 \label{f:rhov200and2200z0}
\end{center}
\end{figure}

\begin{figure}[b]
\begin{center}
 \includegraphics[width=3.4in]{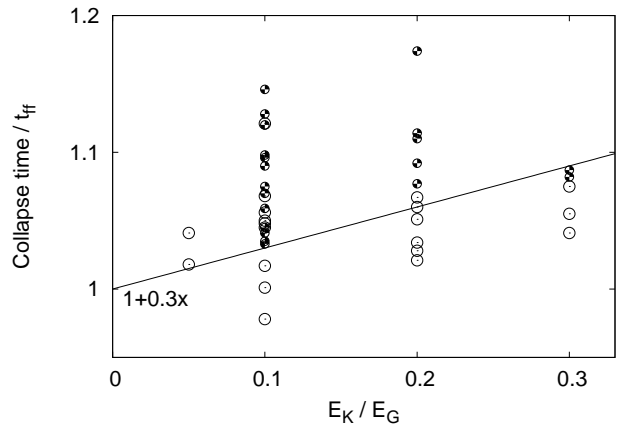}
 \caption{Collapse time versus $E_K/E_G$ ratio in all 40 models. Open circles~-- no
 fragmentation, filled circles~-- yes. Solid line emphasizes that the collapse with
 fragmentation is more delayed than the collapse without fragmentation.}
 \label{f:tclps}
\end{center}
\end{figure}

\begin{figure}[b]
\begin{center}
 \includegraphics[width=3.4in, trim=0.8in 0.25in 0.8in 0.3in, clip]{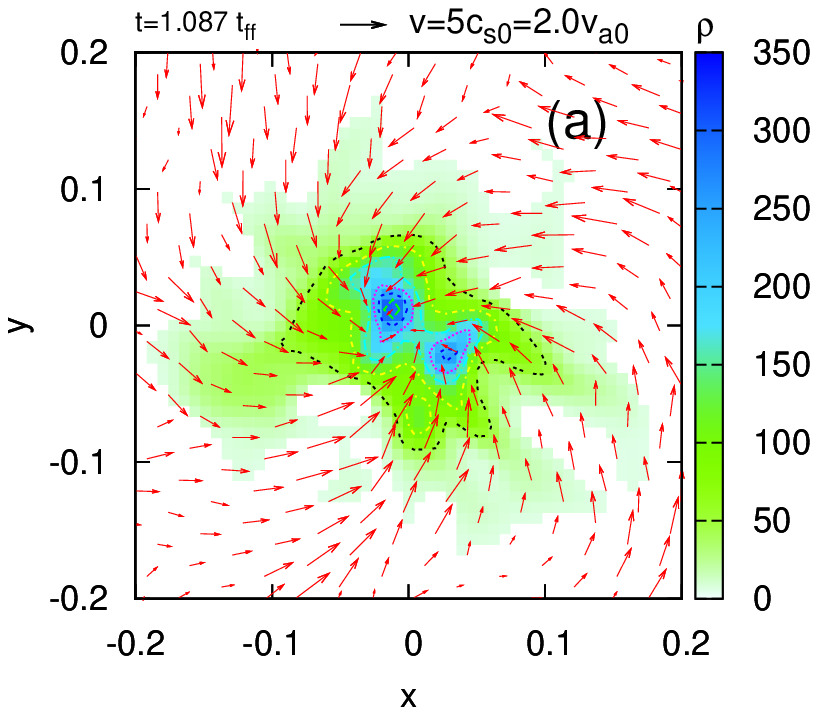}
 \includegraphics[width=3.4in, trim=0.8in 0.25in 0.8in 0.3in, clip]{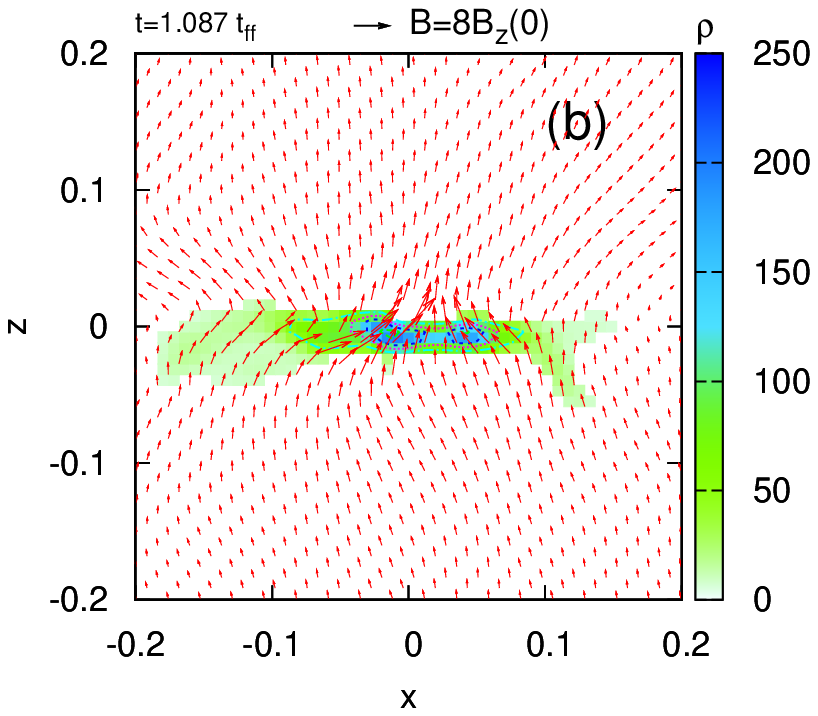} 
 \caption{Final distributions of density and velocity at the plane $z=0$ (a) 
 as well as density and magnetic field at the plane $y=0$ (b) in the central region
 of base model. Only density $\rho>5$ is shown (color scale and contours).}
 \label{f:rhovB_final}
\end{center}
\end{figure}

\begin{figure}[b]
\begin{center}
 \includegraphics[width=3.4in]{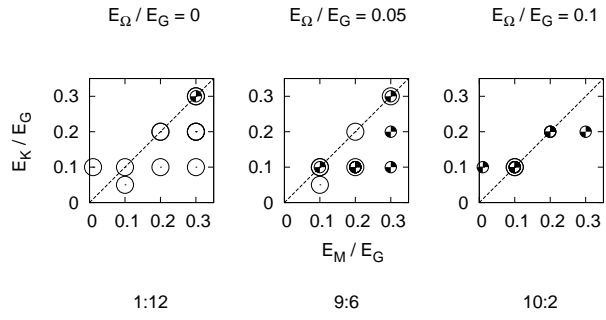}
 \caption{Fragmentation diagram: open circles~-- no, filled circles~-- yes. 
 See yes-to-no ratio at the foot.}
 \label{f:fragDiag}
\end{center}
\end{figure}

\begin{figure}[b]
\begin{center}
 \includegraphics[width=3.4in]{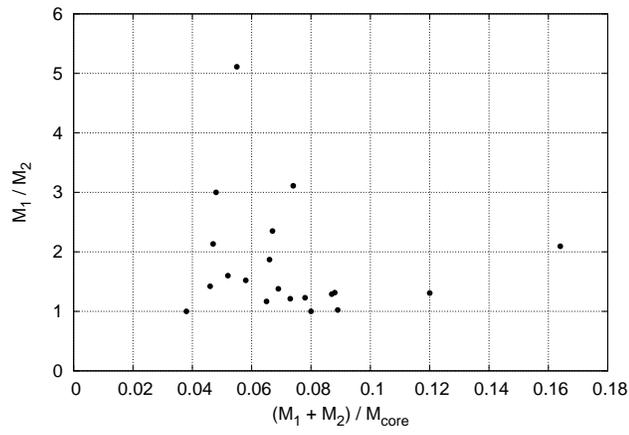}
 \caption{Fragments mass ratio versus their total mass to core mass ratio.}
 \label{f:MsumMrel}
\end{center}
\end{figure}

\begin{figure}[b]
\begin{center}
 \includegraphics[width=3.4in]{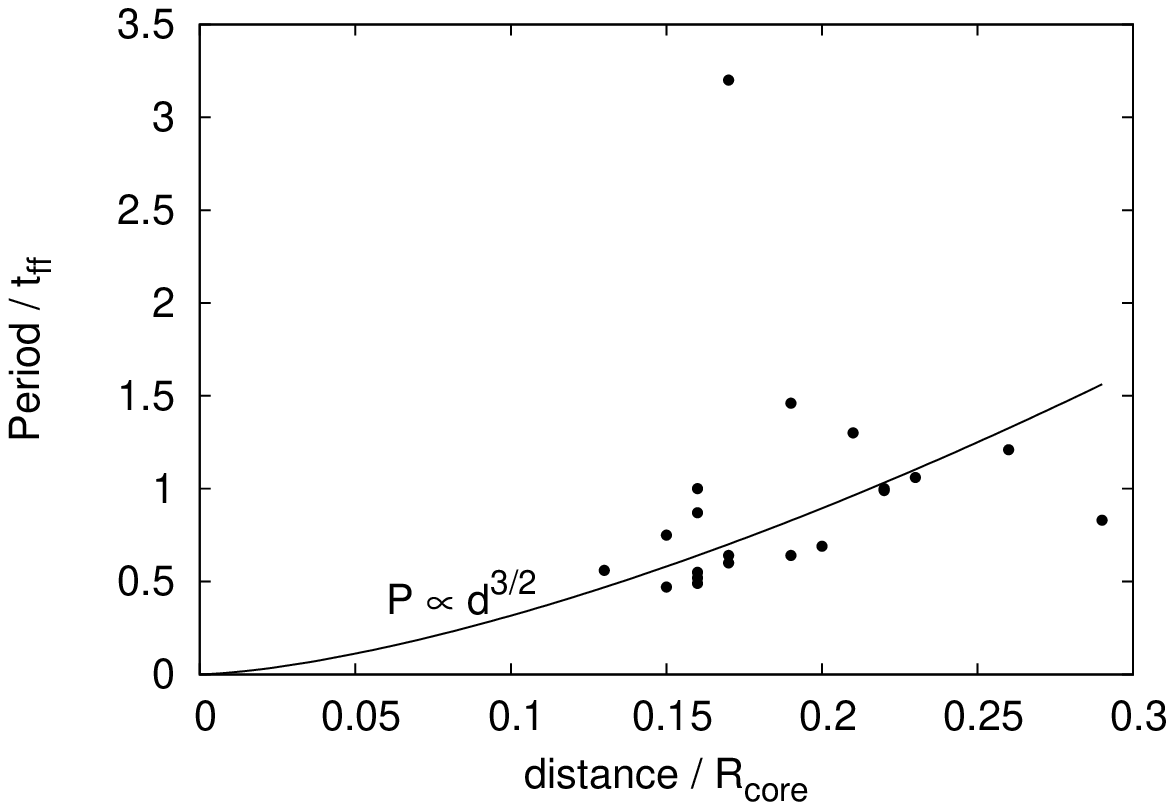} 
 \caption{Orbital period of fragments (in core free fall time units) versus 
 their separation to core radius ratio. The solid curve shows that the correlation
 between distance and period resembles the third Kepler law ($P\propto d^{3/2}$).}
 \label{f:period}
\end{center}
\end{figure}

We have computed successfully 40 models. The basic results are following.
\begin{enumerate}
\item
Turbulence produces density pulsations almost during all the collapse. For
example at fig. \ref{f:rhov200and2200z0} we can see clumpy structure both
at the initial and final simulation stages.
\item
Turbulence usually increases the collapse time by $5-10\%$ (fig. \ref{f:tclps}). 
Larger $\beta$ leads to larger collapse time because of increasing of turbulent 
pressure gradient. The collapse with fragmentation is more delayed than the 
collapse without fragmentation.
\item
During the isothermal collapse stage Jeans mass decreases and some fragments 
become bound. Only two bound fragments can appear when the density increases at 
$100-1000$ times (fig. \ref{f:rhovB_final}a). 
\item
The magnetic field directions in the fragments differ from $\vec B_0$ direction.
For example in fig. \ref{f:rhovB_final}b we can see the declinations of magnetic 
field at $30-45^\circ$ from $z$-axis.
\item
In the absence of rotation the early fragmentation is possible when 
$E_K > E_T$ and $E_K \ge E_M$ (fig. \ref{f:fragDiag} left).
At moderate rotation ($E_\Omega /E_G=0.05$) the early fragmentation becomes 
possible even when $E_K = E_T$ and $E_K < E_M$ (fig. \ref{f:fragDiag} middle).
At faster rotation ($E_\Omega /E_G=0.1$) the early fragmentation occurs 
in most cases (fig. \ref{f:fragDiag} right).
\item
In average, the initial total mass of the bound fragments $\approx 0.07$ of 
core mass, the major to minor mass ratio $\approx 1.5$ (fig. \ref{f:MsumMrel}).
\item
The distance between the mass centres of the fragments is $d=0.13-0.29 R$, 
their orbital period is $P=0.47-3.2 t_{ff}$ (fig. \ref{f:period}). The correlation
between distance and period resembles the third Kepler law ($P\propto d^{3/2}$).
\item
The properties of bound cores almost independent on initial spectral slope 
$\alpha$. This results is surprising and must be checked. 
\item
Since typically $R\simeq 0.01-0.1$ pc then $d\simeq 0.001-0.01$ pc that corresponds 
to the separation in wide binary protostars and stars of the Galaxy field 
(see \cite[Connelley et al. 2008]{Connelley08}).
\end{enumerate}

\section{Conclusions}
We carried out 3D MHD simulations of the collapse and fragmentation 
of rotating turbulent prestellar core embedded into turbulent medium.
We conclude that: 
\begin{itemize}
\item
under realistic parameters only two bound fragments can appear when the core 
density increases at $100-1000$ times, 
\item
the distance between such fragments is about $0.1$ of the initial core radius,
\item
their orbital period is comparable to the initial free fall time of the core, 
\item
these results can explain the origin of binary stars in the Galaxy field 
with semi-major axis in the range $0.001-0.01$ pc.
\end{itemize}

At the next stage of the project we will use the AMR technology realized in the 
code Megalion. That will improve the resolution and lead to more interesting 
results. It is also desirable to develop the turbulent initial conditions without 
MHD discontinuities at the core surface.

\end{document}